\documentclass[aps,prl,preprint,longbibliography,amsmath,amssymb,groupedaddress]{revtex4-2}
\usepackage{graphicx} % for APS, ArXiv
\usepackage{subcaption}

\def\i{i}

\def\vec#1{\mbox{\boldmath $#1$}}
\def\ket#1{\left|#1\right\rangle }
\def\bra#1{\left\langle #1\right|}

\def\refeq#1{Eq.~(\ref{#1})}
\def\fig#1{Fig.~\ref{#1}}

\def\paragraph#1{}
\begin{document}
\title{Theory of Fractionally-magnetized Quantum Ferromagnet}
\author{Isao {Maruyama}}
\email[]{i-maruyama@fit.ac.jp}
\affiliation{Department of Information and Systems Engineering, Fukuoka Institute of Technology, 3-30-1 Wajiro-higashi, Higashi-ku, Fukuoka 811-0295, Japan}
\author{Shin {Miyahara}}
\affiliation{Department of Applied Physics, Fukuoka University, 8-19-1 Nanakuma, Jonan-ku, Fukuoka 814-0180, Japan}
\date{\today}
\begin{abstract}
  We present a theory to realize entangled quantum spin states with fractional magnetization.
  The origin of magnetization reduction is partly emergent antiferromagnetism, that is, spin-liquefaction of ferromagnetism.
  We study a ferromagnetic bilinear coupling region of the spin-\(S\) \(({\geqq} 1)\) bilinear-biquadratic spin chain based on (i) a rigorous eigenstate correspondence between the spin-\(S\) model and spin-\(\frac12\) model and (ii) a numerical exact-diagonalization calculation up to \(S=3\).
  As a result, we obtain a fractional magnetized \(M=1-1/(2S)\) phase, where ground states have quantum entanglement-reflecting corresponding spin-\(\frac12\) antiferromagnetic ground states in a ferromagnetic background.
  This spin-liquefaction theory of ferromagnets can be generalized to any-dimensional lattices even under a magnetic field.
  This fractional ferromagnetism opens the new research field of quantum ferromagnets.
\end{abstract}
\maketitle

\paragraph{Introduction1}
Entangled quantum states have been attracting not only researchers in physics but also developers in quantum computer science.
In condensed-matter physics, antiferromagnets involve many interesting topics, including entangled gapped quantum spin-liquid states \cite{RMP.89.040502} in integer spin-\(S\) chain with a Haldane gap \cite{PRL.50.1153}, and fractionalized \(S/2\) spins that form an entangled spin singlet on a bond in the valence-bond-solid picture of the Affleck--Kennedy--Lieb--Tasaki (AKLT) model \cite{PRL.59.799}.
On the other hand, ferromagnetically ordered states in quantum systems can be approximated as ``classical'' states in the sense that fully polarized local spins have no quantum entanglement.
Is there any ferromagnet with an entangled quantum state?

A key to realizing an entangled ferromagnetic state is to partly create an antiferromagnetic quantum state in a ferromagnetic classical background, that is, ``spin liquefaction'' of a ferromagnet.
When the total spin of a partly emergent ``spin-liquid'' (a phase with nonmagnetic long-range N\'{e}el order) is zero, the coexistent states are fractionally magnetized.
In this Letter, we propose a simple procedure to construct a quantum spin-\(S\) Hamiltonian that leads to the property of a phase transition from fully magnetized ground states to fractionally magnetized ground states under zero magnetic field.
This transition is accomplished by flat-band one-magnon instability and magnetization changes from \(M=1\) to a fraction \(M<1\).
Note that this is not a magnetization-plateau state under an external magnetic field but macroscopically degenerate ferromagnetic ground states with fractional magnetization under zero magnetic field, that is, a ``fractional ferromagnet.''

\begin{figure}
  \includegraphics[width=0.4\textwidth]{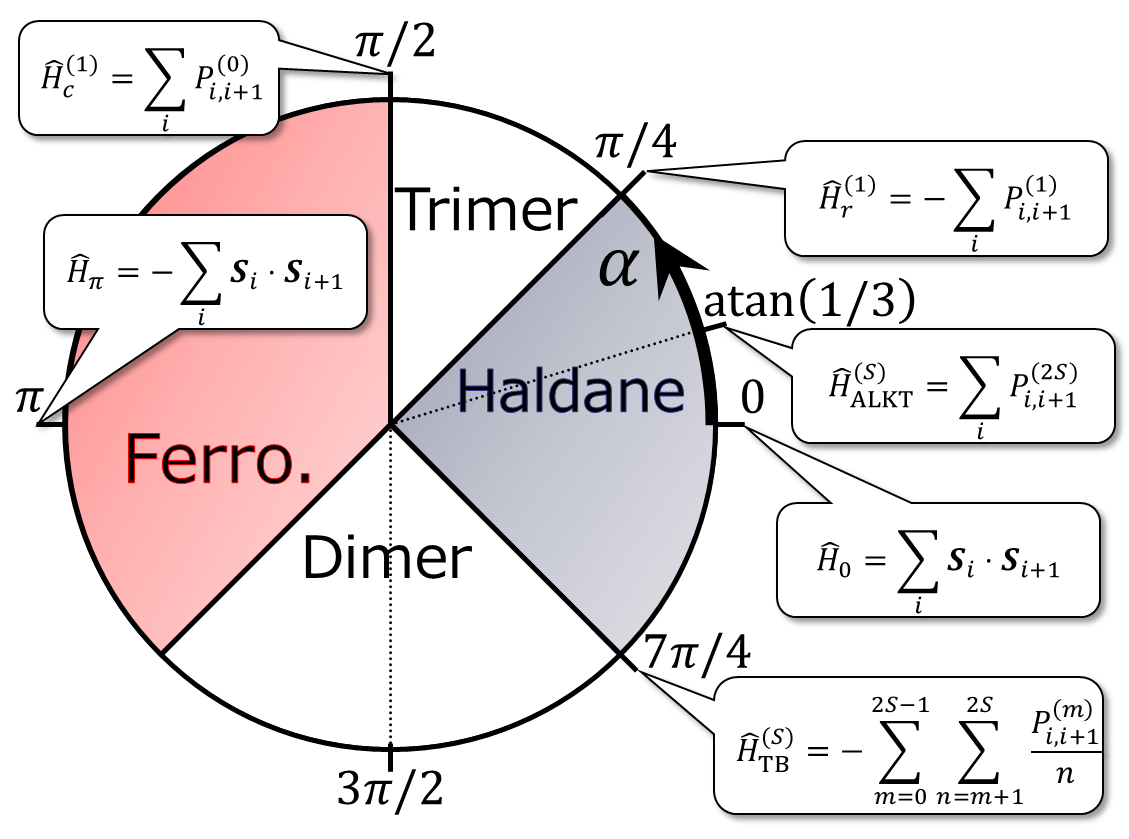}
\caption{Known phase diagram of  \(S=1\) BLBQ chain.
  The high-symmetry points \(\alpha_{r}=\pi/4\) and \(\alpha_c=\pi/2\) are generalized to higher-\( S\)
  in Eqs. (\ref{eq:alphar}) and (\ref{eq:alphac}).
  \(\hat{P}_{ij}^{(s)}\) is a projection operator defined later in \refeq{eq:AAH}.
}
\label{figB}
\end{figure}

\paragraph{Introduction2}
The realization of the spin-liquefaction is supported by a rigorous correspondence between a subset of eigenstates in the spin-\(S\) model and whole eigenstates in the spin-\(\frac12\) antiferromagnetic model.
In other words, the rigorous correspondence is ``eigensystem embedding.''
Thus, it might be interesting even in the context of quantum many-body scars\cite{NP.14.745, PRL.119.030601, PRB.98.235155, PRB.98.235156, PRL.124.180604}.
As an example, we consider a spin-\(S\) (\(S \geqq 1\)) bilinear-biquadratic (BLBQ) chain described by the Hamiltonian
\begin{eqnarray}
  \hat{H}_\alpha^{(S)} = \cos\alpha \sum_{i=1}^N \vec{\hat{S}}_i \cdot \vec{\hat{S}}_{i+1} + \sin\alpha \sum_{i=1}^N \left(\vec{\hat{S}}_i \cdot \vec{\hat{S}}_{i+1}\right)^2
  \label{eq:BLBQ}
\end{eqnarray}
with the periodic boundary condition \(\vec{\hat{S}}_{N+1}=\vec{\hat{S}}_1\).
The phase diagram for the \(S=1\) case, shown in \fig{figB}, has been massively studied \cite{LNP.645.1} and includes the AKLT point at \(\alpha=\arctan{{1\over 3}}\) \cite{PRL.59.799}, the SU(3) point at \(\alpha={\pi\over 4}\) \cite{JETPL.12.225, JMP.15.1675, PRB.12.3795}, and the other high-symmetry points at \(
{5 \pi\over 4}\) \cite{PRB.65.180402}, \(
{3 \pi\over 2}\) \cite{NPB.265.409, PRB.40.4621, JPA.23.15, EPL.9.815}, and \(
{7 \pi\over 4}\) \cite{PLA.87.479, PLA.90.479, NPB.215.317, JPA.21.4397}.
As explained later, for any \(S\), the rigorous eigenstate correspondence
between eigenstates consisting of \(S\) and \(S-1\) spin states in the BLBQ chain and eigenstates in the spin-\(\frac12\) Heisenberg chain (i.e., spin-\(\frac12\) liquefaction) is realized at \(\alpha=\alpha_r\) and \(\alpha=\alpha_r+\pi\), where
\begin{eqnarray}
  \alpha_r=\left\{
  \begin{array}{cc}
  -\arctan\left({1\over 2S(S-2)+1} \right),
  &  S \leqq 3/2   \\
  \pi -\arctan\left({1\over 2S(S-2)+1} \right),
  & S \geqq 2.
  \end{array}
      \right.
  \label{eq:alphar}
\end{eqnarray}
For \(S=1\), \(\alpha_r = {\pi\over 4}\) corresponds to the SU(3) point.
In other words, \(\alpha_r\) is a generalization of the \(S=1\) SU(3) point via preservation of partial SU(2) symmetry for the spin-\(\frac12\) liquefaction.
Note that, because this correspondence at \(\alpha_r\) is for eigenstates, numerical evidence is required to obtain the ground-state properties.

\begin{figure}
\includegraphics[width=0.5\textwidth]{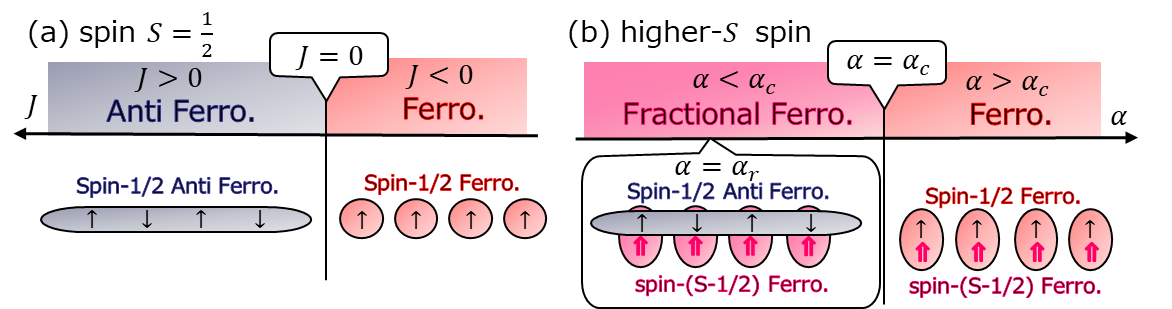}
\caption{(a) Phase transition from ferromagnetic phase to antiferromagnetic phase in \(S=\frac12\) Hamiltonian \(J \sum_{i=1}^N \vec{\hat{s}}_i\cdot \vec{\hat{s}}_{i+1}\). 
  (b) Phase transition at \(\alpha_c\) from ferromagnetic \(M=1\) phase to fractionally magnetized \(M=1-{1\over 2 S}\) phase in the higher-\(S\) BLBQ Hamiltonian \(\hat{H}_\alpha^{(S)}\) described by \refeq{eq:BLBQ}.
  Rigorous ground-state correspondence with spin-\(\frac12\) antiferromagnetic chain realized at \(\alpha_r\) for \(S\geqq 2\).
}
\label{fig:m}
\end{figure}
As a result of numerical calculation of the BLBQ chain, we find that the ground state of the \(S \geqq 2\) BLBQ model at \(\alpha_r\) is equivalent to that of the \(S=\frac12\) antiferromagnetic chain, and the fractionally magnetized state is stabilized in a finite parameter region for \(S \geqq 3/2\). The spin-liquefaction transition from the fully magnetized \(M=1\) phase around the ferromagnetic Heisenberg point \(\alpha=\pi\) to the fractionally magnetized \(M=1-{1\over 2 S}\) phase occurs at
\begin{eqnarray}
  \alpha_c =
  \pi -\arctan\left({1\over 2S(S-1)} \right)
  \label{eq:alphac}
\end{eqnarray}
for \(S \geqq 3/2\), as shown schematically in \fig{fig:m}(b).
This spin-\(\frac12\) liquefaction of the spin-\(S\) system can be considered as a generalization of the ``entire'' spin-liquefaction from the spin-\(\frac12\) ferromagnetic-ordered phase to the antiferromagnetic quantum-disordered phase of the \(S=\frac12\) Hamiltonian \(J \sum_{i=1}^N \vec{\hat{s}}_i\cdot \vec{\hat{s}}_{i+1}\) at \(J=0\), as shown in \fig{fig:m}(a).

\paragraph{projection1}
To explain the theoretical detail, let us start with a spin-projection Hamiltonian of a spin-\(S\) model on any lattice with general coefficients \(J_{ij}^{(s)}\) defined as
\begin{math}
 \hat{H}=\sum_{ij} \sum_{s=0}^{2S} J_{ij}^{(s)} \hat{P}^{(s)}_{ij} 
  ,
\end{math}
where \(\hat{P}_{ij}^{(s)}\) is a projection operator onto the subspace with total spin \(s\in [0,2S]\) for two spins at sites \(i\) and \(j\). 
There is a general relation \cite{PRL.60.531}
\begin{eqnarray}
  \hat{P}_{ij}^{(s)}= \prod_{\substack{n=0 \\ n\neq s}}^{2S} {\vec{\hat{S}}_i\cdot \vec{\hat{S}}_j- q_n \over q_s -q_n}
  ,\;
  (\vec{\hat{S}}_i\cdot \vec{\hat{S}}_j)^n = \sum_{s=0}^{2S} {q_s}^n \hat{P}_{ij}^{(s)}
  ,
  \label{eq:AAH}
\end{eqnarray}
with \(q_s ={s(s+1)/ 2} -S(S+1)\).
Given \(\sum_{s=0}^{2S} \hat{P}_{ij}^{(s)}=1\), the \((2S+1)\)-dimensional parameter space of \(J_{ij}^{(0)}, J_{ij}^{(1)},\ldots,J_{ij}^{(2S)}\) is reduced to \(2S\) dimensions.
By ignoring the positive energy scale factor, the intrinsic parameter space becomes a \(2S\)-dimensional sphere: for \(S=1\), a two-dimensional sphere is a circle parameterized by \(\alpha\), that is, the BLBQ Hamiltonian.
Moreover, the spin-projection Hamiltonian can simply express high-symmetry points of the \(S=1\) BLBQ chain, as summarized in \fig{figB}, by ignoring the positive energy scale factor and energy shift.
In previous studies for \(S=2\),
\(2S=4\) independent parameters are assumed to be \(J_{ij}^{(1)} =J_{ij}^{(3)}=0\) \cite{PRA.85.011601, PRL.114.145301, PRB.105.085140} and \(J_{ij}^{(0)} =J_{ij}^{(1)}=0\) \cite{PRB.83.014409,PRB.81.224430}.
% drop :: and $SO(2S+1)$ extension\cite{PRB.78.094404, PRB.83.060407}.
% drop PRB.84.064409

\paragraph{projection2-1}
In this Letter, we consider the condition \(J_{ij}^{(2S)}=J_{ij}^{(2S-2)}\) for spin-liquefaction, which gives
\begin{eqnarray}
  \hat{H}_r^{(S)}=\left. \sum_{ij} \sum_{s=0}^{2S} J_{ij}^{(s)} \hat{P}^{(s)}_{ij} \right|_{J_{ij}^{(2S)}=J_{ij}^{(2S-2)}},
  \label{eq:r:condition}
\end{eqnarray}
where a subset of eigensystem has a rigorous correspondence with whole eigensystem in the spin-\(\frac12\) Heisenberg model
\begin{math}
  \hat{H}^{(1/2)} = \sum_{ij} (J_{ij}^{(2S)}-J_{ij}^{(2S-1)}) \vec{\hat{s}}_i\cdot \vec{\hat{s}}_j  + \varepsilon_0
\end{math}
with the \(S=\frac12\) operator \(\vec{\hat{s}}_i\) and energy shift $\varepsilon_0=\sum_{ij}(3J_{ij}^{(2S)} +J_{ij}^{(2S-1)})/ 4$.
In short, for any eigenstate \(\ket{\psi}\) of \(\hat{H}^{(1/2)}\), corresponding eigenstates of \(\hat{H}_r^{(S)}\) are rigorously written as \(\ket{\Psi_0}=\hat{C} \ket{\psi}\) with an intertwiner\cite{NPB.338.602,IJMPB.7.3649} \(\hat{C}=\prod_{i=1}^N\left(\ket{S}_i\bra{\uparrow}+\ket{S-1}_i\bra{\downarrow}\right)\), which is a mapping operator from the spin-\(\frac12\) Hilbert space spanned by \(\ket{\uparrow}\) and \(\ket{\downarrow}\), to the spin-\(S\) Hilbert space spaned by \(\ket{S}, \ket{S-1},\ldots, \ket{-S}\).
The degeneracy in \(\hat{H}_r^{(S)}\) is greater than that in \(\hat{H}_r^{(1/2)}\) because of a ferromagnetic moment in \(\ket{\Psi_0}\). The additional degenerate states are \(\ket{\Psi_s}= (\hat{S}_{\mathrm{tot}}^-)^s \ket{\Psi_0}\), where \(\hat{S}_{\mathrm{tot}}^\alpha=\sum_i \hat{S}_i^\alpha\) is a total spin operator. This rigorous eigenstate correspondence is easily proved~\footnote{
  The basic idea is that triplet (singlet) projection of spin-\(\frac12\) model corresponds to spin-\(S\) operator \(\hat{P}^{(2S)}_{ij}+\hat{P}^{(2S-2)}_{ij}\) (\(\hat{P}^{(2S-1)}_{ij}\)).
The proof is based on the following formula
\begin{math}
\left(\hat{P}^{(2S)}_{ij}+\hat{P}^{(2S-2)}_{ij} \right) \ket{T_k}_{ij}=\ket{T_k}_{ij},
  \;\;\;
  \hat{P}^{(2S-1)}_{ij} \ket{S_0}_{ij}=\ket{S_0}_{ij}
     ,
     \left(\hat{P}^{(2S)}_{ij}+\hat{P}^{(2S-2)}_{ij} \right) \ket{S_0}_{ij}=0,
  \;\;\;
  \hat{P}^{(2S-1)}_{ij} \ket{T_k}_{ij}=0
  ,
  \hat{P}^{(s)}_{ij}\ket{T_k}_{ij}=\hat{P}^{(s)}_{ij}\ket{S_0}_{ij}=0,\;\;\; (s\leq 2S-3)
,
\end{math}
for polarized triplet \(\ket{T_1}_{ij}=\ket{S}_i\ket{S}_j=\hat{C}\ket{\uparrow}_i\ket{\uparrow}_j \), \(\ket{T_0}_{ij}=\hat{C}{\ket{\uparrow}_i\ket{\downarrow}_j + \ket{\downarrow}_i\ket{\uparrow}_j\over \sqrt{2}}\), \(\ket{T_{-1}}_{ij}=\hat{C}{\ket{\downarrow}_i\ket{\downarrow}_j}\) and polarized siglet \(\ket{S_0}_{ij}=\hat{C}{\ket{\uparrow}_i\ket{\downarrow}_j - \ket{\downarrow}_i\ket{\uparrow}_j\over \sqrt{2}}\).
The formula is valid not only for \(S\geq 1\) but also for \(S=\frac12\) if we put \(\hat{P}^{(-1)}_{ij}=0\).
Using the above formula, one can prove $\hat{H}_r^{(S)} \hat{C} = \hat{C} \hat{H}^{(1/2)}$.
If $\hat{H}^{(1/2)}\ket{\psi}=\epsilon \ket{\psi}$, one finds $\hat{H}_r^{(S)}\left(\hat{C}\ket{\psi}\right)= \hat{C} \hat{H}^{(1/2)}\ket{\psi}= \epsilon \left(\hat{C}\ket{\psi}\right)$.
More details of the proof and demonstration are provided in \S S.2 in the supplement.
}.
Note also that a numerical calculation is required to confirm that a ground state of \(\hat{H}_r^{(S)}\) may also be written as \(\ket{\Psi_{s}}\).
For eigenstates, however, the correspondence is valid for a general lattice in any dimension, and even under a magnetic field.

\paragraph{projection2-2}
The BLBQ chain, \refeq{eq:BLBQ}, is rewritten as \(\hat{H}_\alpha^{(S)} = \sum_i  \sum_{s=0}^{2S} J_{i i+1}^{(s)} (\alpha) \hat{P}_{i i+1}^{(s)}\), where \(J_{i i+1}^{(s)} (\alpha) = q_s \cos \alpha + {q_s}^2 \sin \alpha\) based on \refeq{eq:AAH}~\footnote{The coefficients $J_{ij}^{(s)}$ as a function of one-parameter $\alpha$ are depicted in \S~S.1 of the supplement.  In addition, $J_{ij}^{(s)}(\alpha_r)$ and  $J_{ij}^{(s)}(\alpha_c)$ are demonstrated for $S=3/2,\ldots, 3$.}.
At the two point $\alpha=\alpha_r$ and $\alpha_{r}+\pi$, given by \refeq{eq:alphar}, the BLBQ chain satisfies the condition \(J_{i i+1}^{(2S)}(\alpha) = J_{i i+1}^{(2S-2)}(\alpha) \).
As a result, a subset of eigensystem in \(H_{\alpha_r}^{(S)}\) correponds to whole eigensystem in spin-\(\frac12\) antiferromagnetic Heisenberg chain.
In addition, \(\hat{H}_\alpha^{(S)}\) at \(\alpha_r\) can be considered as a higher-\(S\) generalization of \(\hat{H}_r^{(1)}=-\sum_{i}\hat{P}^{(1)}_{ii+1}\) at \(\alpha_r={\pi/ 4}\) in \fig{figB}, which leads us to the spin-\(\frac12\) SU(2) model.
This generalization is not the usual SU(\(2S+1\)) generalization with \(J_{ij}^{(2S)}=J_{ij}^{(2S-2)}=\cdots=J_{ij}^{(0)}, J_{ij}^{(2S+1)}=J_{ij}^{(2S-1)}=\cdots=J_{ij}^{(1)}\) \cite{JETPL.12.225, JMP.15.1675, PRB.12.3795, PRL.114.145301}.
% comment on SU(5) for S=2 found in PRL.114.145301

\paragraph{projection-3(new)}
Similarly, as a higher-\(S\) generalization of \(\hat{H}_c^{(1)}=\sum_{i}\hat{P}^{(0)}_{ij}\) at \(\alpha_c\) in \fig{figB}, let us introduce another limitation \(J_{ij}^{(2S)}=J_{ij}^{(2S-1)}\) for the spin-projection Hamiltonian
\begin{eqnarray}
  \hat{H}_c^{(S)}=\left. \sum_{ij} \sum_{s=0}^{2S} J_{ij}^{(s)} \hat{P}^{(s)}_{ij} \right|_{J_{ij}^{(2S)}=J_{ij}^{(2S-1)}< J_{ij}^{(s)}
  , \;(s \leqq 2S-2)}
  \label{eq:pt:condition}
,
\end{eqnarray}
which gives a phase boundary of the ferromagnetic phase (\(J_{ij}^{(2S)}< J_{ij}^{(s)}\)).
For the BLBQ chain $\hat{H}_{\alpha_c}^{(S)}$, at the phase transition point \(\alpha_c\) given by \refeq{eq:alphac}, the condition \(J_{i i+1}^{(2S)}(\alpha)=J_{i i+1}^{(2S-1)}(\alpha)< J_{i i+1}^{(s)}(\alpha)\) is satisfied.
For \(S=\frac12\), this is a quantum phase transition between ferromagnetic and antiferromagnetic phases via a trivial Hamiltonian \(\hat{H}_c^{(1/2)}=0\), as shown in \fig{fig:m}(a).
In general, it is easy to check that the ferromagnetic state \(\ket{0}=\prod_i \ket{S}_i\) and the one-magnon excited state \(\hat{S}^-_{i}\ket{0}\) are ground states of \(\hat{H}_c^{(S)}\) with eigenenergy \(\sum_{ij} J_{ij}^{(2S)}\): that is, the one-magnon flat band degenerate at the ground-state energy.
In addition, other ground states are multi-sublattice N\'{e}el-like states, defined as \(\ket{m}= (\prod_{i\in {\cal L}_m} \hat{S}^-_{i})\ket{0}\) for any \(m\)th sublattice  \(
{\cal L}_m\), where \(\ket{m}\) has \(
{S}_{\text{tot}}^z=N S - N_m\) and \(N_m=|{\cal L}_m|\) is number of \(m\)th sublattice sites.
If a ground state for \(J_{ij}^{(2S-1)}<J_{ij}^{(2S)}\) overlaps with \(\ket{m}\), the ground state has \(S_{\text{tot}}^z=NS - N_m\) and \(S_{\text{tot}}\geqq S_{\text{tot}}^z\), which becomes \(S_{\text{tot}}\geqq S_{\text{tot}}^z=N(S-\frac12)\) for the BLBQ bipartite chain (\(N_m=N/2\)).
It is naively expected that the magnetization jumps to \(M={S_{\text{tot}}/( NS)}={(S-\frac12) / S}\) from \(M=1\) at \(\alpha_c\), whereas numerical evidence is required because other states can be more stable.

\paragraph{gap1:figG}
\begin{figure}
  \begin{subfigure}{0.49\linewidth}
    \includegraphics[width=\linewidth]{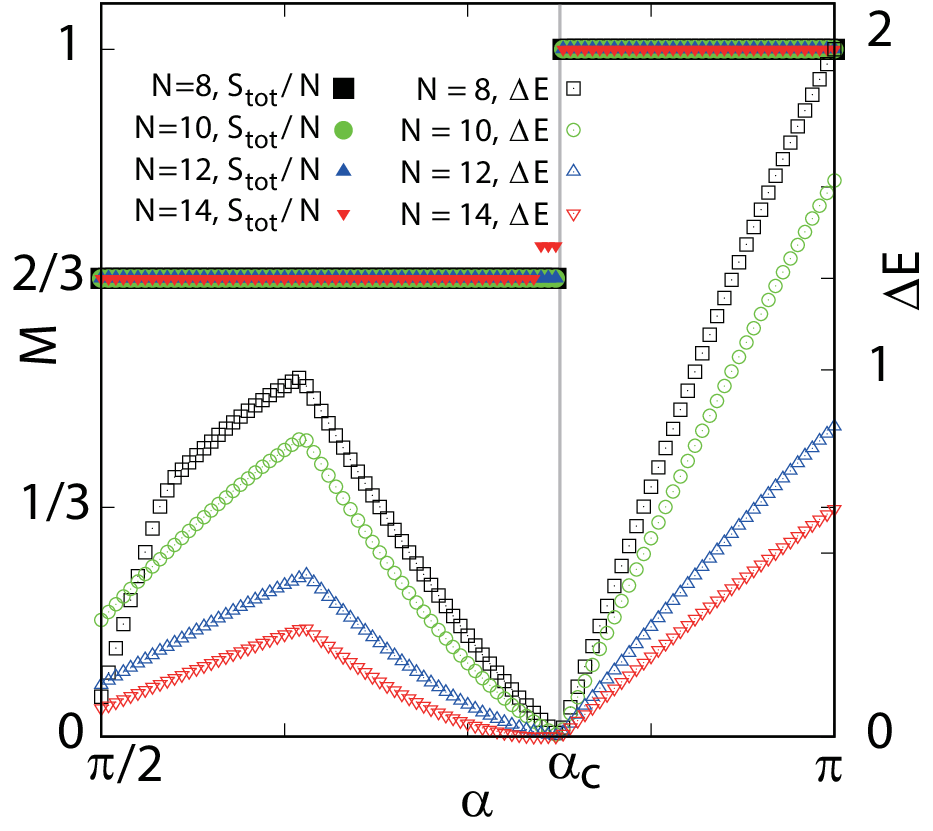} % Arxiv
    \caption{ \(S=3/2\)}
  \end{subfigure}
  \begin{subfigure}{0.49\linewidth}
    \includegraphics[width=\linewidth]{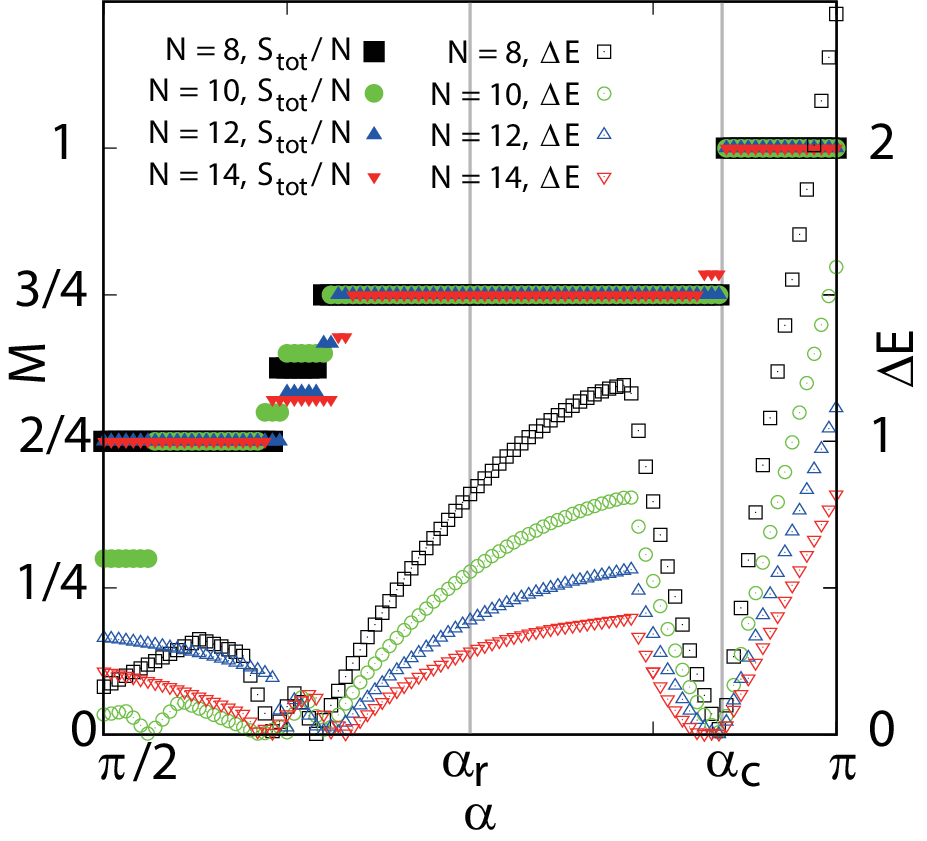}
    \caption{\(S=2\)}
  \end{subfigure}
  \caption{Magnetization \(M={S_{\text{tot}}\over NS}\) and the energy gap \(\Delta E\) for the spin-\(S\) BLBQ \(N\)-site chain Hamiltonian \(\hat{H}_\alpha^{(S)}\) \refeq{eq:BLBQ}
  in the \(S^z_{\text{tot}}=0\) and \(q=0,\pi\) subspace.
  Phase transition from \(M=1\) to \(M=1-{1 /(2S)}\) occurs at \(\alpha_c\) corresponding to \refeq{eq:alphac}.
  (a) \(S=3/2\) and (b) \(S=2\).
}
\label{figG}
\end{figure}
To observe fractional magnetization for the spin-\(S\) BLBQ \(N\)-site chain Hamiltonian \(\hat{H}_\alpha^{(S)}\) [\refeq{eq:BLBQ}], we perform an exact diagonalization with the Lanczos method in the region \(\pi/2 < \alpha <\pi\) up to \(S=3\) by using translational symmetry. Figure \ref{figG} shows the magnetization \(M\) of the ground state and energy gap \(\Delta E\) of the first excited state in the subspace of \(S^z_{\text{tot}}=0\) and wave number \(q=0\) or \(\pi\) for \(S=3/2\) and \(S=2\) and \(N=8, 10, 12\), and \(14\), which shows clear transitions at \(\alpha_c\) [\refeq{eq:alphac}]. 
The magnetization is fractionalized as \(M=1-{1/( 2 S)}\) in a certain region \(\alpha < \alpha_c\).
Here, \(M={S_{\text{tot}}/( N S)}\) is calculated from \(S_{\text{tot}}=f (\langle \vec{\hat{S}}_{\text{tot}}\cdot \vec{\hat{S}}_{\text{tot}} \rangle)\) via \(f (x)={(\sqrt{1+4x}-1)/ 2}\).

\paragraph{gap4:finite-size}
In most of the \(M={(S-\frac12 )/S}\) phases in \fig{figG}, wave-vector \(q\) of the ground state depends on the system-size \(N\) because \(q=0\) (\(\pi\)) for even (odd) \(N/2\), while \(q=0\) for \(\alpha>\alpha_c\).
This even-odd effect of \(N/2\) is consistent with that in the spin-\(\frac12\) Heisenberg chain \footnote{
  Ref.~\cite{PR.128.2131} shown that in spin-\(\frac12\) antiferromagnetic Heisenberg chain \(q=\arg(\langle \hat{T} \rangle)\) and total spin \(S_{\text{tot}}\) becomes ground state has \((q,S_{\text{tot}})=(0,0)\) and excited state has  \((q,S_{\text{tot}})=(\pi,1)\) for even \(N/2\) and ground state has \((q,S_{\text{tot}})=(\pi,0)\) and excited state has  \((q,S_{\text{tot}})=(0,1)\) for odd \(N/2\).
  In the thermodynamic limit, these two states become degenerated, following the fact that energy spectrum \(
{\pi |J| \over 2}|\sin q|\), that is, des Cloizeaux-Pearson mode becomes zero at \(q=0\) and \(\pi\).}.
In detail, in the vicinity of \(\alpha \lesssim \alpha_c\) for large system size \(N \geqq 14\), a state with \(M=1-{1/(2S)}+{1/( N S)}\) has slightly lower energy than that with \(M=1-{1/( 2S)}\), and two states are almost degenerate, which reflects doubly degenerate \(q=0\) and \(q=\pi\) modes in the thermodynamic limit (\(N\rightarrow\infty\))~\cite{Note1}.

\paragraph{gap2:alpha_r}
At the rigorous point \(\alpha_r\), magnetization of the ground states becomes \(M=1-{1/( 2 S)}\) only for \(S\geqq 2\) while \(M\neq 1-{1/( 2 S)}\) for \(S\leqq 3/2\).
A main difference is whether the bilinear term in \refeq{eq:BLBQ} is ferromagnetic (\(S\geqq 2\)) or antiferromagnetic (\(S\leqq 3/2\)).
Since the eigenstate correspondence is rigorous for any \(S\), the eigenstate of spin-\(\frac12\) liquefaction for \(S\leqq 3/2\) can become stable under a magnetic field.
For \(S=1\), the magnetization is \(M=0\) at \(\alpha_r=\pi/4\), which is the critical point between the trimer and the Haldane phase \cite{PRB.74.144426}, as shown in \fig{figB}.
However, a magnetic field induces a phase transition to the magnetized Haldane phase \cite{PRB.83.184433}, which is known to have exact correspondence to the spin-\(\frac12\) model \cite{PRB.58.14709}.
For general \(S\), a rigorous correspondence between the ground state of the BLBQ model and that of the spin-\(\frac12\) antiferromagnetic model can be realized under an external magnetic field.
For \(S=3/2\), magnetic-field-induced spin liquefaction occurs at \(\alpha_r =\arctan(2) \simeq 0.35\pi\). However, this is left as a future problem.

\paragraph{gap3:alpha_c}
The transition point \(\alpha_c\) is at least the phase boundary of fully magnetized ferromagnetic phase \(M=1\).
The proof is simple because ground states and one-magnon excitation are written exactly \footnote{
For \(\alpha>\alpha_c\), the ground-states are \(\left(\hat{S}_{\mathrm{tot}}^-\right)^s\ket{0}\), (\(s\in [0,2NS]\)), with eigenenergy 
\begin{math}
 E_\alpha=NS^2 (\cos\alpha+S^2 \sin\alpha) = \bra{0} \hat{H}_\alpha^{(S)} \ket{0} 
,
\end{math}
which corresponds to classical energy of spin-vector \(\vec{S}_i\cdot \vec{S}_j=S^2 \cos\theta\) for ferromagnetism \(\theta=0\).
One magnon exact excitation has eigenenergy \(E_\alpha+ W_\alpha (1-\cos q)\)  with \(q\) and the hopping element \(W_\alpha = - 2 S \left[\cos\alpha + 2 S (S-1)\sin\alpha\right]\).
Excitation energy is positive due to \(W_\alpha > 0\) for \(\alpha>\alpha_c\).
At \(\alpha_c\), \(W_{\alpha_c}=0\) means one-magnon flat band.
For \(\alpha<\alpha_c\), flipped band is realized due to \(W_{\alpha}<0\) and fully ferromagnetic states become excited states: that is, end of fully magnetized ferromagnetic phase \(M=1\).
}.
As an exact result, the one-magnon band becomes flat at \(\alpha_c\), as is already known from spin-wave theory \cite{PTEP.2014.083101}.
Note that the continuous one-magnon excitation is not depicted in \fig{figG} because the energy gap \(\Delta E\)  is restricted in the sector \(q=0\) and \(\pi\).

\begin{figure}
\includegraphics[width=0.5\textwidth]{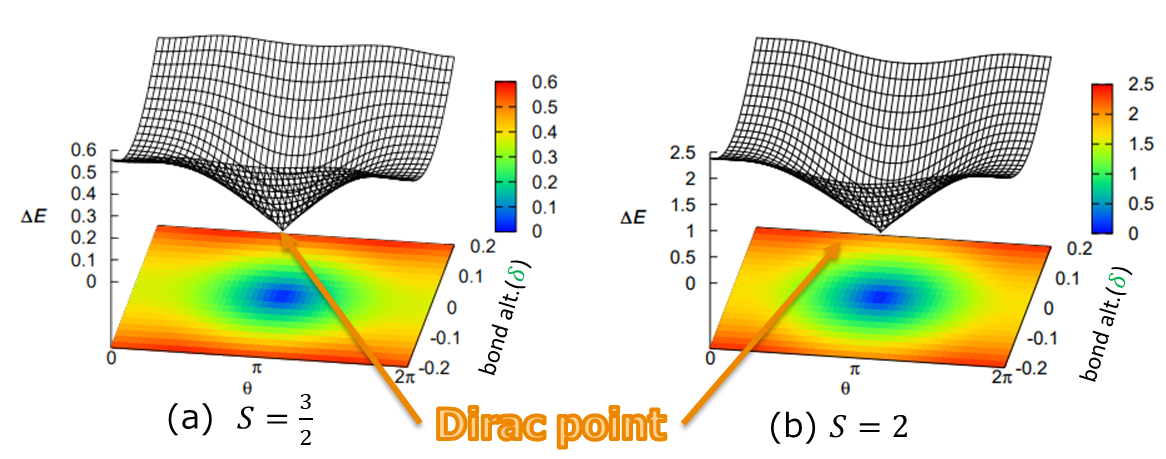}
\caption{Energy gap \(\Delta E\) in the \(S^z_{\text{tot}}=N(S-\frac12)\) sector as a function of phase-twist angle \(\theta\) and bond-alternation \(\delta\) at \(\alpha_0=\alpha_c - 2\pi\times 0.04\) (\(\alpha_r<\alpha_0\lesssim \alpha_c\)) and system-size \(N=8\) for (a)  \(S=3/2\)  and (b)  \(S=2\).
}
\label{figP}
\end{figure}

\paragraph{berry1:figP}
To confirm the thermodynamic limit under the existence of a finite-size gap, we adopt the twisted boundary condition \cite{JPA.30.285} or quantized Berry phase \cite{JPSJ.75.123601}, introducing \(\hat{H}_{\alpha,\delta,\theta}\) with bond-alternation \(\delta\) and boundary twist angle \(\theta\) by using the \(\delta\)-dependent coefficient \(J_{i,i+1}^{(s)}(\alpha,\delta)=[1 +(-1)\delta^i] J_{i,i+1}^{(s)}(\alpha)\) and the \(\theta\)-dependent boundary condition \(\hat{S}^\pm_{N+1}= e^{\pm\i\theta}\hat{S}^\pm_1\) and \(\hat{S}^z_{N+1}=\hat{S}^z_1\).
The energy gap \(\Delta E\) in the sector for \(S^z_{\text{tot}}=N(S-\frac12)\) opens due to finite system size \(N=8\) even in the uniform case (\(\delta=0\)), while the finite gap closes under the twisted boundary condition (\(\theta=\pi\)) only at \(\delta=0\), as shown in \fig{figP} at \(\alpha_{0} = \alpha_c - 2\pi\times 0.04\) (\(\alpha_r<\alpha_0\lesssim \alpha_c\)).

\paragraph{berry2:result}
The result of \fig{figP} is identical to that of dimer singlets in a spin-\(\frac12\) dimerized Heisenberg chain.
In the dimerized limit (\(\delta=1\)), the unique ground state in the subspace for \(S^z_{\text{tot}}=N(S-\frac12)\) is given as a direct-product state of two-site dimer \(\prod_{i=1}^{N/2} (\hat{S}_{2i}^--\hat{S}_{2i+1}^-)\ket{0}\) for \(\alpha<\alpha_c\) exactly \footnote{
  Degenerate eigenstates of two-site Hamiltonian at \(\delta=1\) written by the projections \(\hat{P}^{(s)}_{2i,2i+1}\) are \(\ket{S_{2i,2i+1}=s, S^z_{2i,2i+1}=m}_{2i,2i+1}\) with the eigen energy \(J_{2i,2i+1}^{(s)}\) and \(2s+1\)-fold degeneracy.
  For \(\alpha<\alpha_c\), the minimum coefficient is \(J_{2i,2i+1}^{(2S-1)}\) and the ground states are \(\ket{S_{2i,2i+1}=2S-1, S^z_{2i,2i+1}=m}_{2i,2i+1}\), \((m=-2S+1, m=-2S+2,\ldots, m=2S-1)\) with the energy-gap \(J_{2i,2i+1}^{(2S)}-J_{2i,2i+1}^{(2S-1)}>0\).
  The \(S^z_{\text{tot}}=N(S-\frac12)\) subspace considered in \fig{figP} has unique ground state \(\prod_{i=1}^{N/2} \ket{S_{2i,2i+1}=2S-1, S^z_{2i,2i+1}=2S-1}_{2i,2i+1} =C \prod_{i=1}^{N/2} (\hat{S}_{2i}^- -\hat{S}_{2i+1}^-)\ket{0}\), with normalization constant \(C\).
}.
Twist-angle \(\theta\) dependence appears as \(\hat{S}_{N}^--\hat{S}_{N+1}^-=\hat{S}_{N}^-- e^{-\i\theta}\hat{S}_{1}^-\) in the boundary dimer, while in the other dimerized limit (\(\delta=-1\)) the ground state \(\prod_{i=1}^{N/2} (\hat{S}_{2i-1}^--\hat{S}_{2i}^-)\ket{0}\) does not depend on \(\theta\) due to the absence of the boundary dimer.
This difference of \(\theta\) dependence results in the difference in Berry phase \(\gamma\).
the change of quantized value \(\gamma=0,\pi\) is accompanied by the Dirac cone shown in \fig{figP}.
The two-fold degenerate states at the Dirac point (\(\theta=\pi\)) adiabatically connect to two states separated by the finite-size gap at the periodic boundary condition (\(\theta=0\)).
These two states have \(q=0\) and \(\pi\) for the uniform case \(\delta=0\) depending on the even-odd parity of \(N/2\).
The scenario of the finite-size effect directly corresponds to the \(S=\frac12\) case, which is for the dimer-singlet state \(\ket{\uparrow\downarrow}-\ket{\downarrow\uparrow}=(\hat{S}_{i}^- -\hat{S}_{i+1}^-)\ket{\uparrow\uparrow}=(\hat{S}_{i}^- -\hat{S}_{i+1}^-) \ket{0}_{i,i+1}\) existing in the \(S_{\text{tot}}^z=N(S-\frac12)=0\) subspace; the finite-size gap disappears in the thermodynamic limit \cite{PR.128.2131}.
The Dirac point is observed in most of \(M={(S-\frac12 )/ S}\) phases. However, an interesting discrepancy from the \(S=\frac12\) case occurs in the vicinity of \(\alpha \lesssim \alpha_c\), where the additional Dirac cone appears at \(\theta=0\) and \(\delta=\pm \delta_c\).

\paragraph{fin1:other lattice}
Apart from our numerical results on the chain, the general theory can be applied to previous studies on other lattices.
On a square lattice \cite{PRB.85.140403}, magnetic-field-induced spin-\(\frac12\) liquefaction of the \(S=1\) BLBQ model is realized.
Moreover, on a \(S=1\) BLBQ triangular lattice \cite{PRL.125.057204}, exact correspondence at \(\alpha_r=\pi/4\) exists for \(M\geq 2/3\); for example, the \(M=2/3\)-plateau state must be regarded as the \(1/3\)-plateau state of the spin-\(\frac12\) model and the \(\uparrow \uparrow \downarrow\) state with spin-\(\frac12\) fully polarized in the \(\uparrow \uparrow \uparrow\) background. 

\paragraph{fin2:spin s liquefaction}
Generalizing \(\hat{H}_c^{(S)}\) for the spin-\(\frac12\) liquefaction, it is naively expected that the spin-\(s\) liquefaction point is given by \(J_{ij}^{(2S)}=J_{ij}^{(2S-1)}=\cdots=J_{ij}^{(2S-2s)}<J_{ij}^{(m)}\), \( (m<2S-2s)\) and perturbation from the point toward the other \(2s+1\) parameter space generates several phases, including the ferromagnetic phase (\(M=1\)) and a fractionally magnetized phase (\(M=1-{s\over S}\)).

\paragraph{E1:future theory}
In summary, we present herein the theory of entangled fractionally-magnetized quantum states providing the viewpoint of spin liquefaction on a \(d\)-dimensional lattice.
In the general discussion, the entangled states turn out to be antiferromagnetic entangled states in ferromagnetic background.
To address this fractional ferromagnet, the ferromagnetic region of the spin-\(S\) BLBQ chain was studied numerically.
The fractional magnetization was revealed to have \(M=1-1/(2S)\) even under zero magnetic field; for example, \(M=2/3\) for \(S=3/2\), and \(M=3/4\) for \(S=2\).
Numerous future problems remain.
From a theoretical viewpoint, further calculations (using other numerical or analytical techniques) in the one-dimensional \(S\geqq3/2\) BLBQ model are required to clarify the magnetization curve as a function of external magnetic field, the boundary edge-spin problem (especially under open boundary conditions), the excitation spectrum as a function of \(q\), and the entanglement entropy and spectrum.
A more generic theoretical task is to establish the origin of the interaction in real materials or by optical-lattice experiments.

\paragraph{E2:Ferri}
The spin-\(\frac12\) liquefaction at \(\alpha_r\) opens up further discussion, for example, a comparison with ferrimagnetism \cite{AP.12.137}.
In a ferrimagnet, spin-\(s\) and the \(S\) Hamiltonian break one-site translation symmetry because \(s\neq S\), whereas a fractional ferromagnet holds that symmetry.
The difference can induce anomalous low-energy excitations in the BLBQ model. In particular, the fractional ferromagnet at \(\alpha_r\) exhibits linear magnon excitation, which reflects the two-fold degeneracy of N\'{e}el-like states in the uniform Hamiltonian (i.e., the des Cloizeaux--Pearson mode \cite{PR.128.2131}), and its existence is guaranteed thanks to the rigorous correspondence to spin-\(\frac12\) antiferromagnetic chain.

\paragraph{E3:fin}
As mentioned above, fractional ferromagnets are not conventional ferrimagnets.
In addition, the fractional ferromagnetic state is not the classical ferromagnetic state near the quantum critical point \cite{RMP.88.025006}.
Even after spontaneous magnetization, the ground state of a fractional ferromagnet has quantum entanglement corresponding to the spin-\(\frac12\) antiferromagnetic state.
For the quantum entanglement in a fractional ferromagnet, the external magnetic field has the potential to be a tool to manipulate an entangled quantum state, which can be useful in the context of quantum computer science.
From the viewpoint of condensed-matter physics, the key word ``quantum magnet'' has been used and accepted for antiferromagnets.
Given that the present theory abolishes the prejudice that ferromagnetism is classical, quantum magnets will also be used for fractional ferromagnets.

\paragraph{E4:last}
To summarize, this Letter develops the new frontier of quantum spin states (i.e., ``quantum ferromagnet''), which opens new field not only in fundamental physics but also in quantum computer science.
\\

\begin{acknowledgments}
The authors thank Hosho Katsura for stimulating discussions.
This work was supported by Japan Society for the Promotion of Science (JSPS) KAKENHI Grants No. 22H01171.
The computation was partly carried out using the computer resource offered by Research Institute for Information Technology, Kyushu University.
\end{acknowledgments}

%\bibliography{../macro,../wiki}
%\input{draft.bbl.save.1.1}

\end{document}